\documentclass[a4paper]{jpconf}
\usepackage{graphicx}

\newcommand{\hfbax}{\sc hfb-ax}

\newcommand{\madnesshfb}{\sc madness-hfb}
\newcommand{\rr} {\mathbf{r}}
\newcommand{\bea}{\begin{eqnarray}}
\newcommand{\eea}{\end{eqnarray}}

\begin{document}
\title{Coordinate-Space Hartree-Fock-Bogoliubov Solvers for Superfluid Fermi Systems in Large Boxes}

\author{J C Pei$^{1,2}$, G I Fann$^{3}$, R J Harrison$^{3,4}$,  W  Nazarewicz$^{1,2,5}$, J  Hill$^{3}$, D  Galindo$^{3}$, J Jia$^{3}$}

\address{$^1$Department of Physics and Astronomy, University of Tennessee, Knoxville, TN 37996, USA}
\address{$^2$Physics Division, Oak Ridge National Laboratory, P.O. Box 2008, Oak Ridge, TN 37831, USA}
\address{$^3$Computer Science and Mathematics Division, Oak Ridge National Laboratory, Oak
Ridge, TN 37830, USA}
\address{$^4$Department of Chemistry, University of Tennessee, Knoxville, TN 37996 USA}
\address{$^5$Institute of Theoretical Physics, Warsaw University,  ul. Ho\.za 69, 00-681 Warsaw, Poland}

\ead{peij@ornl.gov}

\begin{abstract}
The self-consistent Hartree-Fock-Bogoliubov problem in large boxes can
be solved accurately in the coordinate space with the recently
developed solvers {\hfbax} (2D) and {\madnesshfb} (3D). This is essential
for the description of superfluid Fermi systems with complicated topologies and significant
spatial extend, such as fissioning nuclei, weakly-bound nuclei, nuclear matter in the neutron star rust,  and ultracold Fermi atoms in elongated traps.  The {\hfbax}
solver based on B-spline techniques uses a hybrid MPI and
OpenMP programming model for parallel computation for distributed
parallel computation, within a node multi-threaded LAPACK and BLAS
libraries are used to further enable parallel calculations of large
eigensystems. The {\madnesshfb} solver uses a novel
multi-resolution analysis based adaptive pseudo-spectral techniques to
enable fully parallel 3D calculations of very large systems. In this
work we present benchmark results for {\hfbax} and  {\madnesshfb}  on ultracold
trapped fermions.

\end{abstract}

\section{Introduction}

The Hartree-Fock-Bogoliubov (HFB) equation of the Density Functional Theory (DFT) is suitable for describing superfluid Fermi systems by
properly accounting for the self-consistent coupling between the particle-hole and particle-particle mean fields. Recently, we solved HFB equations in complicated geometries to describe  nuclei and ultra-cold polarized
Fermi gases~\cite{jcpei}. The general HFB equation for a polarized system
can be written as:
\begin{equation}\label{HFBm}
\begin{array}{c}
  \left(
\begin{array}{cc}
h_a(\textbf{r})-\lambda_a& \Delta(\textbf{r}) \vspace{3pt} \\
\Delta^{*}(\textbf{r}) &-h_b(\textbf{r})+\lambda_b\\
\end{array}
\right)\left(
\begin{array}{c}
u(\textbf{r}) \vspace{2pt}\\
v(\textbf{r}) \\
\end{array}
\right)=E\left(
\begin{array}{c}
u(\textbf{r}) \vspace{2pt}\\
v(\textbf{r}) \\
\end{array}
\right),
\end{array} \nonumber
\end{equation}
where $h_a$ and $h_b$ are Hartree-Fock Hamiltonians for the two
spin components, $\lambda_a$ and $\lambda_b$ are the corresponding
chemical potentials, and $\Delta$ is the pairing potential.

In some cases, solving the HFB equation in large boxes in coordinate space   is
essential. Weakly-bound nuclei and the large-amplitude nuclear collective motion such as fission and fusion are examples of  problems that require large-box
calculations~\cite{hfbax}. The ultra-cold atoms trapped in elongated
optical traps also require a description involving large spatial dimensions \cite{jcpei}, as well as
HFB description of nucleonic pasta phases in the neutron star crusts \cite{newton}.  All these problems are both interesting and important but
the underlying calculations are challenging.

Many HFB solvers used in the nuclear physics context are based on the basis expansion method employing harmonic oscillator wave functions. The configuration-space  method is efficient but offers a fairly poor accuracy for cases involving  weakly-bound systems and large deformations, see discussion in Ref.~\cite{hfbax}. On the other hand, solving HFB equations directly in coordinate-space can offer very precise results.
Unfortunately, because of numerical challenges involved, HFB calculations in non-spherical geometries require high computation cost. In this context,
multi-core processor architectures
such as the Jaguar and Kraken Cray XT5  with 12 cores per node, or Hopper Cray XE6 supercomputer with  24 cores per node,
promise to revolutionize the deformed HFB problem. To fully take advantage of unique computational capabilities,  HFB
solvers should be adapted and improved in terms of scalability.

To this end, we developed two parallel HFB codes: 2D {\hfbax} for
axially symmetric systems and {\madnesshfb} for fully 3D systems. Both solvers
take advantage  of modern multi-core architectures. The following  sections, overview computational and
numerical techniques employed by  these two codes and present
benchmark calculations  for ultra-cold Fermi atoms. To illustrate  the deformed HFB problem in coordinate space, we present an extreme application
to  novel pairing  phases in
polarized cold Fermi gases in extremely elongated traps~\cite{pei10}.

\section{Two-dimensional HFB solver {\hfbax}}

In {\hfbax}~\cite{hfbax}, the wave functions are
discretized on  a 2-D grid  ($r_{\alpha}$, $z_{\beta}$) with the
$M$-order B-splines:
\begin{equation}
\psi_{n\Omega^\pi }(r_{\alpha},
z_{\beta})=\displaystyle
\sum_{i,j}B^M_i(r_{\alpha})B^M_j(z_{\beta})C_{n\Omega^\pi
}^{ij}, \nonumber
\end{equation}
where the coefficients $C^{ij}_n$ can be obtained by  diagonalizing the Hamiltonian matrix.
%%%%%%
\begin{figure}[hbdg]
\center
\includegraphics[width=0.8\textwidth]{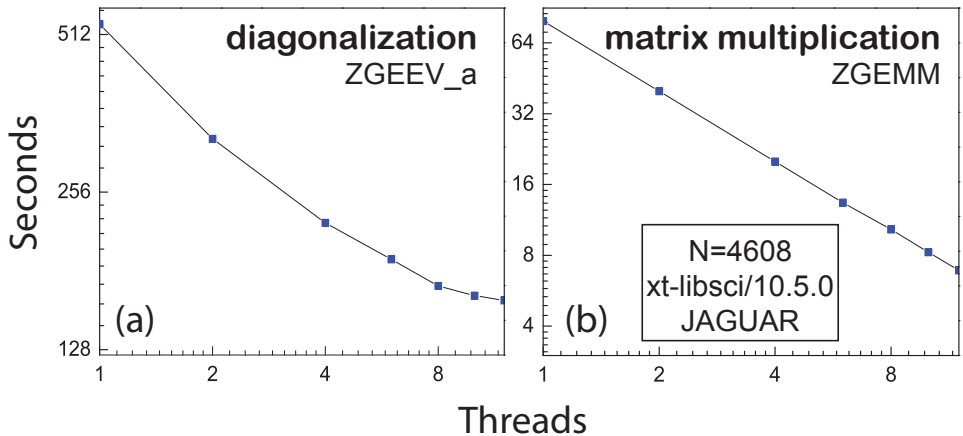}
\caption{\label{threads} Left: performance of the complex eigensolver  ZGEEV{\_}a using
  different number of threads (ZGEEV\_a is a modified version of
  LAPACK's ZGEEV). Right: performance of the complex
  matrix multiplication routine ZGEMM. The two routines are employed by {\hfbax} and tested on  Jaguar with the xt-libsci/10.5.0 library.}
\end{figure}
%%%%%
To reduce computational costs, we assume reflection symmetry; hence, the intrinsic parity $\pi$ is a good
quantum number that can be used to enumerate HFB eigenstates. Another quantum number preserved by {\hfbax}  is the angular momentum projection $\Omega$  on the symmetric axis.
The boundary conditions are  implemented through  first- and second-order derivative operators.
The  precision of computations depends on the  mesh size,  order of B-splines, and the box size.  Those parameters determine the numerical effort involved.

In {\hfbax}, the diagonalization of the complex Hamiltonian matrix takes the bulk of the
computing time.  Diagonalization blocks of given  $\Omega$ and
$\pi$ quantum numbers are assigned to computing cores through
MPI communications.  We modified the LAPACK diagonalization
routine ZGEEV to calculate only the selected eigenvectors.  For
calculations of a heavy nucleus with the  mesh size of 0.6 fm,
10-th order B-splines, and a box size of 24 fm, the rank of Hamilton
matrix is $N$=4608. To take the advantage of multi-core architectures,
we use the threaded LAPACK library, which can speed up the
diagonalization by a factor of 3 with 6 threads, as shown in
Fig.~\ref{threads}(a).
For the matrix multiplication problem  in Fig.~\ref{threads}(b), the routine ZGEMM scales perfectly.  For other matrix
operations, we use OpenMP for parallel
calculations. The modified Broyden
method is  used to accelerate the iteration convergence~\cite{hfbax}.  The
hybrid MPI and OpenMP programming have made {\hfbax} an accurate and fast HFB
solver for large box calculations. Besides the threaded library on
Cray machines, one can also utilize the threaded AMD's ACML library or the
threaded Intel MKL library. In the near future, the multi-core+GPU Cray
XT6 machine is expected, and the corresponding MAGMA
library~\cite{magma} can further improve the capability of {\hfbax}.

The precision of {\hfbax} has been demonstrated in calculations of
weakly-bound nuclei in which the coupling to scattering continuum is
essential~\cite{hfbax,pei11}.  In large-box calculations, very dense
quasi-particle spectrum is obtained, and the non-resonant continuum contribution can
be precisely taken into account by means of the direct integration \cite{pei11}.  Furthermore, {\hfbax} is
expected to provide precise solutions for the continuum QRPA
description of excited states by means of the recently developed  finite-amplitude method \cite{Sto11}.  The recent success of {\hfbax}
is the prediction  of  rapidly oscillating pairing potential in highly
elongated traps \cite{pei10} by  using  the SLDA~\cite{Bulgac07}
and ASLDA~\cite{Bulgac08} energy density functionals for polarized Fermi
gases.

\section{3D {\madnesshfb} Solver}

The software {\madnesshfb} \cite{fann,fann1} is a fast $O(N log \epsilon)$ method based
on the multiresolution analysis and low separation rank methods of
representing and approximating functions and operators for solving
Schr\"odinger  and Lippman-Schwinger equations up to user-determined precision
$\epsilon$.  MADNESS is an acronym for Multiresolution Adaptive
Numerical Environment for Scientific Simulations \cite{madness}. The application of
multiresolution analysis separates the behavior of functions and
operators at different length scales in a functional way.
Interesting mathematical and computational feature in our MADNESS
library implementation is that each of the operators and
wave functions has its own adaptive structure of refinement to achieve
and guarantee the desired accuracy.

\begin{figure}[htb]
\center
\includegraphics[width=0.45\columnwidth]{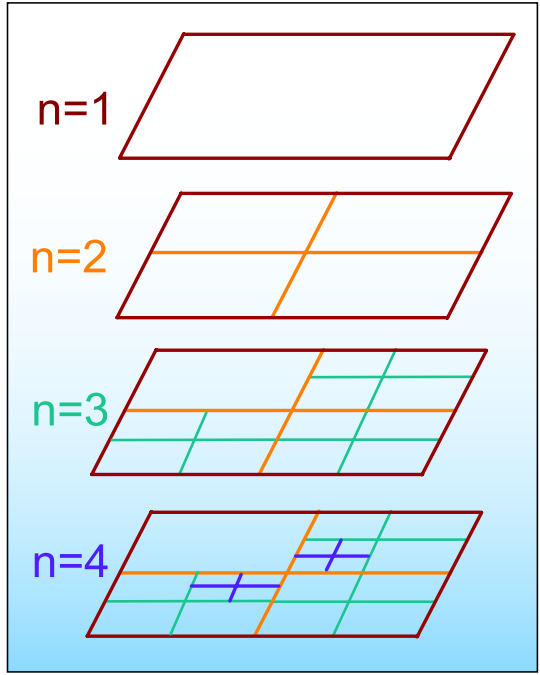}
\caption{\label{multi} A schematic picture illustrating the two-scale generated refinement process of a multi-resolution representation of a function in 2D with 3-levels.
 }
\end{figure}

The representation of functions and operators in MADNESS is based on
an adaptive pseudo-spectral representations of functions and operators
using the discontinuous Alpert's multiwavelets \cite{Alpert,ABGV} and the
low-separation rank representations and applications of Green's
functions.  Alpert's multiwavelets have compact support.  For representing
functions in 3D, we use tensor products of 1D multiwavelets.  The
two-scale relations provides for an adaptive approximation of the
expansion in terms of multiwavelet basis as a function of accuracy in
terms of the regularity of the solutions.  The support of the bases
functions in each level resembles that of structure adaptive refinement. A schematic figure of such a refinement
process is shown in Fig.~\ref{multi}.
%%%%
Using  two-scale relations, the multiwavelet basis
provides a way of constructing an adaptive pseudo-spectral method.
The multiwavelet basis is a local basis of degree $k$ defined on the
interval $(0,1)$ generated from scaling functions which are defined by
shifted, rescaled and orthonormalized Legendre polynomials from degree
0 to $k-1$ from $(-1,1)$ with a value of 0 defined outside of $(0,1)$.
The subspace, of $L_2((0,1))$, spanned by scaling functions of degree
$k$ is denoted by $V^k = { \phi_k (x) =\sqrt{k+1/2} P_i (2x-1) \ on
  \ (0,1) }$ where $P_i (x)$ is the $i$-th Legendre polynomial on
$(-1,1)$.  For each level $n$, we further define an ascending sequence
of the subspaces $V_n^k = { \phi_{jl}^n (x) = 2^{n/2} \phi_{j}(2^n x -
  l), j=0, ...., k-1,\ l=0, ..., 2^n -1 }$.  Specifically, $V_{k}^0
\subset V_{k}^1$; the multiwavelets basis, denoted by $W_{k}^{0}$, is
the orthogonal complement of $ V_{k}^{0}$ in $V_{k}^{1}$.  For each
level $n$, we can further construct $W_{k}^n$ in the same way of
shifting and rescaling the basis functions of $W_{k}^{0}$ as in
$V_{k}^{n}$ to construct $W_{k}^{n}$.  Thus,

\begin{equation}
\begin{array}{l}
  V_0^k \subset V_1^k \subset V_2^k \subset ... \subset V_{n}^k,  \vspace{5pt} \\
  V_n^k = V_0^k + W_{0}^k + W_1^k+...+ W_{n-1}^k . \nonumber
  \end{array}
\end{equation}

The notable features of the multiwavelet bases are that at each
level the support of the basis functions are of width $2^{-n}$, and
that there is an exact algebraic relationship between the basis
functions at level $n$ and level $n+1$.  In particular, the
coefficients of a representation of a smooth function in the
multiwavelet basis decays proportionally to the width of the
interval at level $n$. Thus, we can estimate the accuracy of the
representation and truncate coefficients below a specified threshold
for a sparse representation.

In 3D, we use a tensor product basis generated from the 1D
multiwavelet basis; the representation of operators uses a
non-standard approximation or a low separation rank approximation
based on Gaussians.  The structure of the support of a function
resembles that of an oct-tree.  Each node of the oct-tree consists of
a tensor of coefficients.  The truncation of the sparse coefficients
produce a pruned tree. The oct-tree is distributed across the nodes of
a massively parallel computer referenced by a global hash-table for
each node of the tree.
The singular integral operators are represented using a low-separation rank
approximation of the Green's functions \cite{BM} with details described in \cite{HFB}.

The MADNESS library uses a combined MPI and pthread parallel computing
method in a task based computing model with a task graph scheduling and
queue on each node.  On each node of a parallel multicore computer,
one core is devoted to processing internode communication via MPI, and  one
core is devoted to handling thread scheduling and task allocation and
queue scheduling within the node.  The code is polymorphic with the
use of C++ templates.  Each operation on functions and the application
of operators are  defined in terms of tasks to be performed on the
nodes or tensors of the ``oct-trees'' in the representation the
functions or operators.  In addition, a data path and flow path
dependency analysis is performed to determine when different steps of
the code can be overlapped and scheduled in each node's task queue so
that distributed multi-threading computation can be performed.
Furthermore, since different functions and operations have different
data and work-loads, a user directed dynamic load-balancing of data
memory can be performed to permit more efficient data and work load
distribution.

The solution methodology for the {\madnesshfb} consists of two main
steps: (i) diagonalization of the Hamiltonian matrix using wave functions expanded in
the multiwavelet bases in 3D,  and (ii) computation of each of the wave functions
using the associated Lippman-Schwinger equation.  Recall that the Green's function for the operator $-\Delta - \lambda$ is $e^{-\lambda r}/r$, the Yukawa potential. The solving procedure is similar to that for solving Hartree-Fock problems~\cite{fann}. In short, the solution algorithm is:
\begin{itemize}
\item Obtain a set of guess wave-function $u_{0}$ in the multiwavelet basis.
\item Iterate until convergence:
\begin{itemize}
\item Compute  and diagonalize the Hamiltonian matrix ${\cal H}$ to obtain the latest orthogonal wave functions;
\item Compute densities, properties and gradients;
\item Compute the approximation of the Yukawa potential for each eigenvalue $\lambda$;
\item Solve the Lippman-Schwinger equation by convolving with the Green's function based on Yukawa potential, $u_{n+1} = G \star V u_{n}$.
\end{itemize}
\item Compute observables.
\end{itemize}

We carried out benchmark calculations with {\hfbax} and {\madnesshfb} for 100 ultra-cold fermionic atoms in elongated traps. The cold Fermions at the unitary limit can be described by the superfluid density functional SLDA~\cite{Bulgac07}. The single-particle Hamiltonian of SLDA can be written as:
\begin{equation}\label{slda}
h ({\bf r})= -\frac{\alpha\hbar^2\nabla^2}{2m}+\frac{\beta(3\pi^2\rho({\bf r}))^{2/3}}{2}
   -\frac{|\Delta({\bf r})|^2}{3\gamma \rho^{2/3}({\bf r})}+V_{ext}({\bf r}),
\end{equation}
where the deformed external trap potential is:
\begin{equation}\label{vtrap}
V_{ext}({\bf
r}) = V_{0}\left[1-\exp\left(-\frac{\omega^2(
x^2+y^2+z^2/\eta^2)}{2V_{0}}\right)\right].
\end{equation}
The densities of spin-up ($\rho_\uparrow$) and spin-down
($\rho_\downarrow$) atoms, the pairing densities $\kappa$,
and pairing gaps $\Delta$ can be expressed in terms of the HFB two-component eigenvectors (\ref{HFBm}):
\begin{eqnarray}\label{dens}
\rho_{\uparrow}(\rr)& =& \sum_i f_i|u_i(\rr)|^2,
\hspace{18pt} \rho_{\downarrow}(\rr) = \sum_i (1-f_i)|v_i(\rr)|^2, \nonumber\\
\kappa(\rr)& =& \sum_{i} f_iu_i(\rr)v_i^*(\rr),
\hspace{5pt} \Delta(\rr) =  -g_{eff}(\rr)\kappa(\rr),
\end{eqnarray}
where $\rho=\rho_\uparrow + \rho_\downarrow$ and  $g_{eff}(\rr)$ is the regularized pairing strength. In
Eq.~(\ref{dens}), $f_i=[1+\exp(E_i/kT)]^{-1}$ and the temperature is  $kT$=0.01. The parameters $\alpha$, $\beta$ and $\gamma$ in Eq.(\ref{slda}) are taken as 1.14, $-$0.553 and $-$1/0.0906, respectively \cite{Bulgac07}.
We work in trap units for which $\hbar=m=\omega=1$. The trap aspect ratio $\eta$ in (\ref{vtrap}) denotes the trap elongation.  In experiments, the adopted optical trap is highly elongated with $\eta$ up to 50~\cite{Par06a}.
The highly elongated trap is interesting because it provides a connection
to quasi one-dimensional systems. In {\madnesshfb}, we employed very large boxes: $(x, y, z)$[-100, 100] for $\eta=5$ and [-160, 160] for $\eta=16$. In {\hfbax}, the 2D box sizes are 10.5$\times$35 ($\eta=5$) and 9.1$\times$70 ($\eta=16$).  The potential depth $V_0$ is 12 and 10 for $\eta=5$ and $\eta=16$, respectively.

\begin{figure}[htb]
\center
\includegraphics[width=0.8\textwidth]{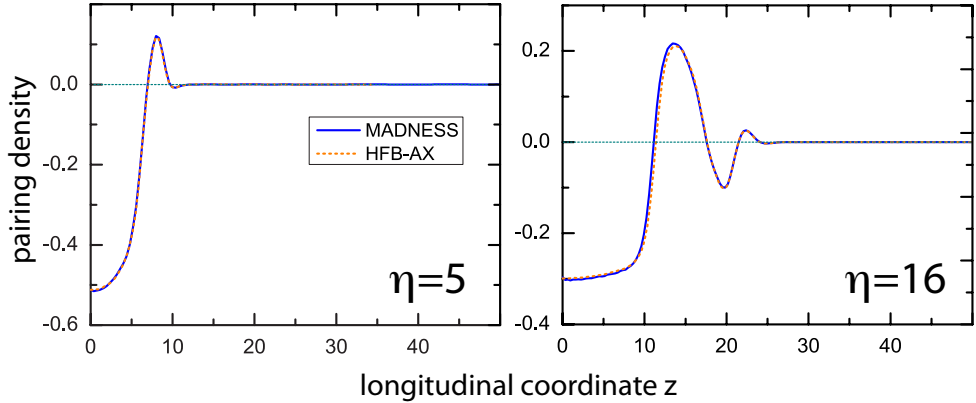}
\caption{\label{compare} Pairing densities $\kappa(x=0, y=0, z)$ calculated by {\madnesshfb} and {\hfbax} for 100 polarized atoms in an elongated trap with $\eta=5$ (left) and $\eta=16$ (right).}
\end{figure}
\begin{figure}[htb]
\center
\includegraphics[width=0.8\textwidth]{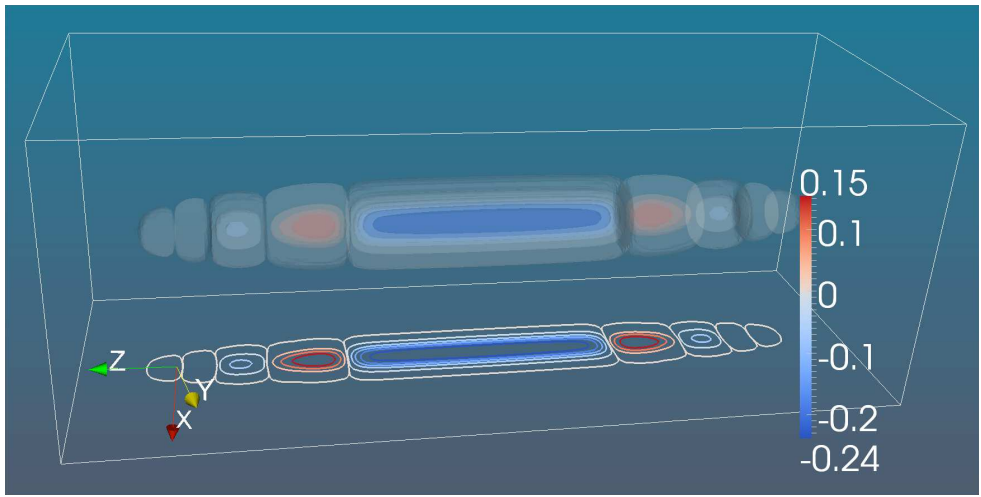}
\caption{\label{3Dview} Pairing density  $\kappa(x, y, z)$, calculated by {\madnesshfb} for the elongated trap with  $\eta=16$. The box scale in view is $x$[-12, 12], $y$[-12, 12] and $z$[-32, 32].
 }
\end{figure}

%%%%
Figure~\ref{compare} displays the calculated pairing densities $\kappa(x=0, y=0, z)$ for ultra-cold Fermi gas  with polarization of 0.2.
The energy cutoff is taken as 6.75 and 5.05 for $\eta=5$ and 16, respectively.
It is seen that the two solvers agree very well at $\eta$=5. At $\eta=16$, there appears  a very small difference between HFB solutions;
this may be an indication that an even larger box is needed in {\hfbax}.
It is to be noted that at $\eta=16$ there appear  coexisting solutions which are close in energy \cite{baksmaty,pei10}.
Figure~\ref{3Dview} shows the 3D  pairing density distribution computed by {\madnesshfb} for  very elongated system with $\eta=16$. The characteristic transversal oscillations of the pairing field are indicative of the  Larkin-Ovchinnikov  phase \cite{pei10}.
%%%%%

In general, the computational effort of  3D HFB calculations  is significantly greater than in the 2D case. Thus, the adaptive representation using
multiwavelet techniques and   parallel computing  are  crucial for the success of {\madnesshfb}. To put things in perspective,  {\madnesshfb} takes about 5 minutes per
iteration on 5000 cores for  $\eta=5$ and $\eta=16$,
while {\hfbax} takes about 10 minutes per iteration for $\eta=5$
 on 400 cores and 25 minutes for $\eta=16$ on 800 cores. From this trend, we conclude that  the parallel
performance of {\madnesshfb} is superior to {\hfbax} in large-box
calculations.

\section{Conclusions}

The  coordinate-space HFB solvers   {\hfbax} and
{\madnesshfb} have been developed for large-box calculations of
superfluid Fermi systems. The 2D solver {\hfbax} has been adapted to the
multi-core supercomputers by using a hybrid MPI and OpenMP programming
model. The 3D solver {\madnesshfb} is based on multi-resolution multiwavelet
techniques and more sophisticated hybrid MPI and pthreads based
task parallelism programming methodologies. A scaling benchmark test for 100
ultracold Fermions in elongated traps has been carried out.  Developments are underway to efficiently use the next
generation of multi-core+GPU architectures for more demanding nuclear  problems.

\ack

This work was supported in part by the U.S. Department of Energy under
Contract Nos.  DE-FG02-96ER40963 (University of Tennessee) and
DE-FC02-09ER41583 (UNEDF SciDAC Collaboration).  It is
also partially sponsored by the Office of Advanced Scientific Computing
Research; U.S. Department of Energy. The work was partially performed
at the Oak Ridge National Laboratory, which is managed by UT-Battelle,
LLC under Contract No. De-AC05-00OR22725.

Computational resources were provided through an INCITE award
``Computational Nuclear Structure" by the National Center for
Computational Sciences (NCCS) and National Institute for Computational
Sciences (NICS) at Oak Ridge National Laboratory, and the National
Energy Research Scientific Computing Center (NERSC).

\end{document}